\documentclass[12pt]{article}
\usepackage{epsfig}
\usepackage{a4,isolatin1}
\usepackage{amsmath,amsfonts,latexsym, amssymb}

\newtheorem{satz}{Theorem}[section]
\newtheorem{defi}[satz]{Definition}

\newtheorem{bem}[satz]{Remark}
\newtheorem{lemma}[satz]{Lemma}

\newtheorem{bsp}[satz]{Example}

\newtheorem{conclusion}[satz]{Conclusion}
\newtheorem{ob}[satz]{Observation}

\newtheorem{res}[satz]{R\'esum\'e}

\newcommand{\mcal}{\mathcal}

\newcommand{\tit}{\textit}

\newcommand{\bewende}{$ \hfill \Box $}

\begin{document}
\thispagestyle{empty}
\begin{center}

{\LARGE{\bf  (Quantum) Space-Time as a\\ Statistical Geometry
 of Fuzzy Lumps\\ and the Connection with\\ Random Metric Spaces}}

\vskip 1cm

{\bf Manfred Requardt}\\{(email:
\small requardt@theorie.physik.uni-goettingen.de)\\
Institut f\"ur Theoretische Physik \\
Universit\"at G\"ottingen \\
Bunsenstrasse 9 \\
37073 G\"ottingen \quad Germany}
\vskip 0.3cm
and\\
\vskip 0.3cm
{\bf Sisir Roy}\\
{(email:{\small sisir@isical.ac.in ) \\
Indian Statistical Institute \\
Calcutta -700 035 \\
India}}
\end{center}

\vspace{0.5 cm}

\begin{abstract}
  We develop a kind of \tit{pregeometry} consisting of a web of
  overlapping \tit{fuzzy lumps} which interact with each other. The
  individual lumps are understood as certain closely entangled
  subgraphs (\tit{cliques}) in a dynamically evolving network which,
  in a certain approximation, can be visualized as a time-dependent
  \tit{random graph}. This strand of ideas is merged with another one,
  deriving from ideas, developed some time ago by Menger et al, that
  is, the concept of \tit{probabilistic-} or \tit{random metric
    spaces}, representing a natural extension of the metrical
  continuum into a more microscopic regime. It is our general goal to
  find a better adapted geometric environment for the description of
  microphysics. In this sense one may it also view as a dynamical
  randomisation of the \tit{causal-set} framework developed by e.g.
  Sorkin et al. In doing this we incorporate, as a perhaps new aspect,
  various concepts from \tit{fuzzy set theory}.\\[0.3cm]

\end{abstract} \newpage
\setcounter{page}{1}
\section{Introduction}
There exists a certain suspicion among the scientific community
that nature may be discrete or rather ``behaves discretely'' on the
Planck scale. But even if one is willing to agree with this ``working
philosophy'', it is far from being evident what this vague metaphor
actually means or how it should be implemented into a
concrete and systematic inquiry concerning physics and mathematics in the
Planck regime.

There are basically two overall attitudes as to ``discreteness on the
Planck scale'', the one  starts (to a
greater or lesser degree) from continuum concepts (or more
specifically: concepts being more or less openly inspired by them) and
then tries to detect or create modes of ``discrete behavior'' on very
fine scales, typically by imposing quantum theory in full or in parts
upon the model system or framework under discussion. We call this the ``top
down'' approach.

There are prominent and very promising candidates in this class like
e.g.\tit {string theory} or \tit{loop quantum gravity} with e.g.
\tit{Spin networks} emanating from the latter approach.  As these
approaches are widely known we refrain from citing from the
vast corresponding literature. We recommend instead two more recent
reviews dealing with the latter approach but containing also some remarks
about the former one (\cite{Smolin} and
\cite{Rovelli}). As a beautiful introduction to the conceptual
problems of quantum gravity in general may serve e.g. \cite{Isham}.

In the following investigation we undertake to describe how
macroscopic space-time (or rather, its underlying \tit{mesoscopic} or
\tit{microscopic} substratum) is supposed to emerge as a
\tit{superstructure} of a \tit{web of lumps} in a fluctuating
dynamical cellular network. One may call this the ``bottom up'' approach.
In doing this, two strands of research are joined, which, originally,
started from different directions. The one is the \tit{cellular
  network} and \tit{random graph} approach, developed by one of the
authors (M.R.), the other the \tit{statistical geometry of lumps}, a
notion originally coined by Menger and coworkers and being further
developed by various groups (see e.g.
\cite{Me},\cite{Rosen},\cite{Roy},\cite{Schweizer1}).  It is worth
mentioning that Einstein himself was not against such a grainy
substratum underlying our space-time continuum (see the essay of
Stachel in \cite{Einstein}).

The point where these different strands meet is the following. In the
dynamical network approach of (M.R.) macroscopic space-time is
considered to be a coarse-grained emergent phenomenon (called an
\tit{orderparameter manifold} in \cite{5}). It is assumed to be the
result of some kind of \tit{geometric phase transition} (very much in
the spirit of the physics of self-organisation). This framework was
developed in quite some detail in e.g. \cite{5}. We argued there that,
what we consider to be the elementary building blocks of continuous
space-time, i.e. the so-called \tit{physical points}, are on a finer
scale actually densely entangled subclusters of \tit{nodes} and
\tit{bonds} of the underlying network or graph. In \cite{5} we called
them also \tit{cliques} (which denote in graph theory the maximal
complete subgraphs or maximal subsimplices of a given graph).

We further argued there that the substructure of our space-time
manifold consists in fact of two stories, the primordial network,
dubbed by us $QX$, and, overlying it, the web of lumps or cliques,
denoted by $ST$, which can also be viewed as a coarser mesoscopic
kind of network with the cliques or lumps as \tit{supernodes} and with
\tit{superbonds} which connect lumps having a non-void overlap. This
correrspondence raises the possibility to relate the lumps or cliques
of \cite{5} with the lumps occurring in the approach of Menger et al.

One should however note that the two philosophies are not entirely the
same. In \cite{5} and related work the lumps emerge from a more
primordial discrete dynamical substratum and consequently have a
specific internal structure. In the approach of Menger et al (at least
as far as we can see) they figure as the not further resolvable
building blocks of space-time if one approaches the so-called
Planck-regime from above, i.e. from the continuum side. In other
words, the former approach is more bottom-up oriented while the latter
one is more top-down. About such not further resovable scales of
space-time (where the ordinary continuum picture ends) was of course
also speculated by quite a few other people, most notably Wheeler (see
e.g. \cite{Wheel}, p.1203 ff.).

Our personal working philosophy is that space-time at the very bottom
(i.e. near or below the notorious Planck scale) resembles or can be
modeled as an evolving information processing cellular network,
consisting of elementary modules (with, typically, simple internal
discrete state spaces) interacting with each other via dynamical bonds
which transfer the elementary pieces of information among the nodes.
That is, the approach shares the combinatorial point of view in
fundamental space-time physics which has been advocated by e.g.
Penrose. It is a crucial and perhaps characteristic extra ingredient
of our framework that the bonds (i.e. the elementary interactions) are
not simply dynamical degrees of freedom (as with the nodes their
internal state spaces are assumed to be simple) but can a fortiori,
depending on the state of the local network environment, be switched
on or off, i.e. can temporarily be active or inactive! This special
ingredient of the dynamics hopefully allows the network to perform
\tit{geometric phase transitions} into a new ordered phase displaying
a certain \tit{two-story structure} to be explained below. This
conjectured emergent geometric order can be viewed as kind of a
discrete \tit{proto space-time} or \tit{pregeometry} carrying
metrical, causal and dimensional structures (as to the \tit{cellular
  network approach} cf. \cite{1} to \cite{5}, the other point of view
is expounded in e.g. the book of Roy; \cite{Roy} to which we also
refer for a more complete list of references).

We will see that several types of distance concepts do emerge in our
analysis on the various scales of resolution of space-time. On the in
our scheme most fundamental level, that is, the primordial network (or
graph), we have a natural distance concept (node distance). The same
holds for the network of lumps, derived from it (the \tit{clique
  graph}; see below). If we leave these fundamental levels (of
relatively discrete behavior) and enter the realm of quasi-continuity,
other possibilities do emerge. It is interesting to relate these
various concepts to each other and to distance concepts discussed e.g.
in \cite{Schweizer1} or \cite{Menger2}. We show in particular that our
network of fuzzy lumps leads naturally to the concept of \tit{random
  metric spaces} (see section (\ref{metric})). Spaces of this kind
should emerge at the interface between the presumed discrete Planck
scale scenario and a perhaps \tit{quasi-continuous}, that is,
coarse-grained spatial environment, connecting this primordial level
with the ordinary continuous regime.

Our overall working philosophy is that geometry has to emerge from a
purely \tit{relational} picture. We hence embark on a reconstruction
of the continuum concepts of ordinary physics and/or mathematics from
more primordial \tit{pregeometric} (and basically discrete) ones
which, in our view, hold sway on the Planck scale. We presume that
this may also shed some light on the better understanding of quantum
theory as such and on the various top-down approaches mentioned above.
We note in passing, that this point of view has a venerable history of
its own, beginning with e.g.  Leibniz. A perhaps related but
technically different approach was developed by Isham et al some time
ago (see e.g.  \cite{I1} or \cite{I2}), who chose to quantise metrical
concepts (a more recent approach is \cite{Topos}). It would be an
interesting task to relate these two approaches to each other.

As another precursor may be considered the \tit{causal-set} approach
of Sorkin et al (see e.g. \cite{Bombelli} or \cite{Sorkin}). In a
sense our framework adds dynamics, viz. interaction, and random
behavior to this picture.

\section{The Cellular Network Environment}
Motivated by the above working philosophy, we emulate the underlying
substratum of our world, or, more specifically, of our space-time
(quantum) vacuum (containing however in addition all the existing
quantum and macro objects as extended excitation patterns!) by what we
call a cellular network.  This discrete structure consists of
elementary \tit{nodes}, $n_i$, which interact (or exchange
information) with each other via \tit{bonds}, $b_{ik}$, playing the
role of (in this context) not further reducible elementary interactions.
The possible internal structure of the nodes (modules) or bonds
(interaction chanels) is emulated by discrete internal state spaces
carried by the nodes/bonds. The node set is assumed to be large but
finite or countable. The bond $b_{ik}$ is assumed to connect the nodes
$n_i,n_k$. The internal states of the nodes/bonds are denoted by
$s_i$, $J_{ik}$ respectively. As our philosophy is, to generate
complex behavior out of simple models we, typically, make simple
choices for them, one being e.g.
\begin{equation} s_i \in q\cdot \mathbb{Z}\quad,\quad J_{ik}\in
  \{-1,0,+1\} \end{equation}
with $q$ an elementary quantum of information.

As in our approach the bond states are dynamical degrees of freedom
which, a fortiori, can be switched off or on, the \tit{wiring}, that
is the pure \tit{geometry} of the network is also an emergent, dynamical
property and is \tit{not} given in advance. Consequently, the nodes
and bonds are typically not arranged in a more or less regular array, a
lattice say, with a fixed nea-/far-order. This implies that
\tit{geometry} will become to some extent a \tit{relational} (Machian)
concept and is no longer an \tit{ideal element} (cf. the more detailed
discussion in \cite{Quantum}, which deals primarily with the emergence
of quantum theory as a consequence of the geometric fine structure of
such a network).

On the other side, as in cellular automata, the node and bond states
are updated (for convenience) in discrete clock time steps,
$t=z\cdot\tau$, $z\in\mathbb{Z}$ and $\tau$ being an elementary clock
time interval. This updating is given by some \tit{local} dynamical
law (examples given below). In this context \tit{local} means that the
node/bond states are changed at each clock time step according to a
prescription with input the overall state of a certain neighborhood
(in some topology) of the node/bond under discussion.  We want however
to emphasize that $t$ is \tit{not} to be confounded with some
\tit{physical time}, which, for its part, is also considered to be an
emergent coarse grained quantity. The well known \tit{problem of time}
is, for the time being, not treated in detail in the following, as it
presents a big problem of its own, needing a careful and separate
analysis (see e.g. \cite{time} or \cite{Butter}). That is, at the
moment the above clock time is neither considered to be dynamical nor
observer dependent. We discussed however the presumed emergence of a
new primordial time scale which sets the scale for the regime where
quantum fluctuations hold sway in \cite{Quantum}.

A simple example of such a local dynamical law we are having in mind
is given in the following definition.
\begin{defi}[Example of a Local Law]
At each clock time step a certain {\em quantum} $q$
is exchanged between, say, the nodes $n_i$, $n_k$, connected by the
bond $b_{ik}$ such that
\begin{equation} s_i(t+\tau)-s_i(t)=q\cdot\sum_k
  J_{ki}(t)\end{equation}
(i.e. if $J_{ki}=+1 $ a quantum $q$ flows from $n_k$ to $n_i$ etc.)\\
The second part of the law describes the {\em back reaction} on the bonds
(and is, typically, more subtle). This is the place where the
so-called `{\em hysteresis interval}' enters the stage. We assume the
existence of two `{\em critical parameters}'
$0\leq\lambda_1\leq\lambda_2$ with:
\begin{equation} J_{ik}(t+\tau)=0\quad\mbox{if}\quad
  |s_i(t)-s_k(t)|=:|s_{ik}(t)|>\lambda_2\end{equation}
\begin{equation} J_{ik}(t+\tau)=\pm1\quad\mbox{if}\quad 0<\pm
  s_{ik}(t)<\lambda_1\end{equation}
with the special proviso that
\begin{equation} J_{ik}(t+\tau)=J_{ik}(t)\quad\mbox{if}\quad s_{ik}(t)=0
\end{equation}
On the other side
\begin{equation} J_{ik}(t+\tau)= \left\{\begin{array}{ll}
\pm1 & \quad J_{ik}(t)\neq 0 \\
0    & \quad J_{ik}(t)=0
\end{array} \right. \quad\mbox{if}\quad
\lambda_1\leq\pm
  s_{ik}(t)\leq\lambda_2
\end{equation}
In other words, bonds are switched off if local spatial charge
fluctuations are too large, switched on again if they are too
small, their orientation following the sign of local charge
differences, or remain inactive.

Another interesting law arises if one exchanges the role of
$\lambda_1$ and $\lambda_2$ in the above law, that is, bonds are
switched off if the local node fluctuations are too small and are
switched on again if they exceed $\lambda_2$.  We emulated all these
laws on a computer and studied a lot of network properties. The latter
law has the peculiar feature that it turned out to have very short
\tit{transients} in the simulations, i.e. it reaches an
\tit{attractor} in a very short clock time. Furthermore these
attractors or state-cycles turned out to be very regular, that is, they
had a very short period of typically six, that is, the whole network returned
in a previous state after only six clock time steps, which is quite
remarkable, given the seeming complexity of the evolution and the huge
phase space (\cite{Nowotny}).
\end{defi}
Remarks:\hfill\\
\begin{enumerate}
\item It is important that, generically, such laws, as introduced
  above, do not lead to a reversible time evolution, i.e. there will
  typically exist \tit{attractors} or \tit{state-cycles} in total
  phase space (the overall configuration space of the node and bond
  states). On the other hand, there exist strategies (in the context
  of cellular automata!)
to design particular \tit{reversible} network laws (cf. e.g. \cite{12}) which are,
however, typically of second order. Usually the existence of attractors
is considered to be important for \tit{pattern formation}. On the
other side, it may suffice that the phase space, occupied by the
system, shrinks in the course of evolution, that is, that one has a
flow into smaller subvolumes of phase space.
\item In the above class of laws a direct bond-bond interaction is not
  yet implemented. We are prepared to incorporate such a (possibly
  important) contribution in a next step if it turns out to be
  necessary. In any case there are not so many ways to do this in a
  sensible way. Stated differently, the class of possible physically
  sensible interactions is perhaps not so numerous.
\item As in the definition of evolution laws of \tit{spin
  networks} by e.g. Markopoulou, Smolin and Borissov (see \cite{13} or
\cite{14}), there are in our case more or less two possibilities:
treating evolution laws within an integrated space-time formalism or
regard the network as representing space alone with the time evolution
being implanted via some extra principle ( which is the way we have
chosen above). The interrelation of these various approaches and
frameworks, while being very interesting, is however  far from obvious
at the moment and needs a separate detailed investigation.
\end{enumerate}
\begin{ob}[Gauge Invariance] The above dynamical law depends nowhere on the
absolute values of the node charges but only on their relative
differences. By the same token, charge is nowhere created or destroyed. We have
\begin{equation}\Delta(\sum_{QX}s(n))=0\end{equation}
To avoid artificial ambiguities we can e.g. choose a fixed reference
level and take as initial
condition respectively constraint
\begin{equation}\sum_{QX}s(n)=0\end{equation}
\end{ob}

There are many different aspects of our class of cellular networks one
can study in this context. One can e.g. regard them as complex
dynamical systems, or one can undertake to develop a statistical or
stochastic framework etc. In a purely geometric sense, however, they
are evolving \tit{graphs}. As we are in this paper primarily concerned with the
analysis of the \tit{microstructure} of (quantum) space-time, it seems
to be a sensible strategy to supress, at least in a first step, all
the other features like e.g.  the details of the internal state spaces
of nodes and bonds and concentrate instead on their pure \tit{wiring
  diagram} and its \tit{reduced (graph) dynamics}. This is already an
interesting characteristic of the network (perhaps somewhat
reminiscent of the \tit{Poincar\'e map} in the theory of chaotic
systems) as bonds can be switched on and off in the course of clock
time so that already the wiring diagram will constantly change.
Furthermore, as we will see, it encodes the complete \tit{near-} and
\tit{far-order structure} of the network, that is, it tells us which
regions are experienced as near by or far away (in a variety of
possible physical ways such as strength of correlations or with
respect to some other physically meaningful metric like e.g.  \tit{statistical
  distance} etc.).  Evidently this is one of the crucial features we
expect from something like physical space-time. In
the above simple scenario with $J_{ik}=\pm 1 $ or $0$ one can e.g.
draw a \tit{directed bond}, $d_{ik}$, if $J_{ik}=+1$, with
$J_{ik}=-J_{ki}$ implied, and delete the bond if $J_{ik}=0$. This
leads to a (clock) time dependent graph, $G(t)$, or \tit{wiring
  diagram}. In other words, we will deal in the following mainly with
the evolution and structure of large \tit{dynamical} graphs.

We close this section with a brief r\'esum\'{e} of the characteristics
an interesting network dynamics should encode (in our view).
\begin{res}
  Irrespectively of the technical details of the dynamical evolution
  law under discussion it should emulate the following, in our view
  crucial, principles, in order to match certain fundamental
  requirements concerning the capability of {\em emergent} and {\em
    complex} behavior.
\begin{enumerate}
\item As is the case with, say, gauge theory or general relativity,
  our evolution law on the surmised primordial level should implement
  the mutual interaction of two fundamental substructures, put
  sloppily: ``{\em geometry}'' acting on ``{\em matter}'' and vice
  versa, where in our context ``{\em geometry}'' is assumed to
  correspond in a loose sense with the local and/or global bond states
  and ``{\em matter}'' with the structure of the node states.
\item By the same token the alluded {\em selfreferential} dynamical
  circuitry of mutual interactions is expected to favor a kind of
  {\em undulating behavior} or {\em selfexcitation} above a return
  to some uninteresting `{\em equilibrium state}' as is frequently
  the case in systems consisting of a single component which directly
  acts back on itself. This propensity for the `{\em autonomous}'
  generation of undulation patterns is in our view an essential
  prerequisite for some form of ``{\em protoquantum behavior}'' we
  hope to recover on some coarse grained and less primordial level of
  the network dynamics.
\item In the same sense we expect the overall pattern of switched-on and
 -off bonds to generate a kind of ``{\em protogravity}''.
\end{enumerate}
\end{res}
\section{The Network of Lumps}
In \cite{5} we argued that our microscopic cellular network, $QX$, may
be capable of performing a \tit{geometric phase transition} into a
more ordered phase, dubbed $QX/ST$, with elementary building blocks the
maximal connected subgraphs or \tit{cliques} (maximal subsimplices)
occurring in the primordial graph, belonging to $QX$. We further
argued that this emergent \tit{orderparameter manifold} (as it plays a
role similar to ordinary orderparameters in the statistical mechanics
of phase transitions) may constitute a stochastic protoform of our
ordinary space-time or quantum vacuum with the cliques as protoforms
of physical points (lumps), having an inner dynamical structure and
being entangled with the other (overlapping) lumps via common nodes or
bonds.

The stochastic aspects are brought in by the underlying dynamical
network law, which induces, among other things, a certain amount of
creation and annihilation of bonds among the microscopic nodes. As a
consequence the size and shape of the cliques or lumps fluctuates in
the course of network evolution. These aspects have been analyzed in
quite some detail in \cite{5} within the framework of \tit{random
  graphs}.

This derived coarser network, viz. the \tit{clique graph} or \tit{web
  of lumps}, is defined as follows. The cliques are represented by new
\tit{meta-nodes}, the respective \tit{meta-bonds} represent overlap of
cliques (in form of common nodes).
\begin{ob}Note that, while this new network may be regarded as being
  coarser in some sense, it may nevertheless in general consist of
  much more nodes and bonds than the underlying primordial network.
  Usually there are much more maximal subsimplices than primordial
  nodes, as a given node will typically belong to quite a few
  different maximal subsimplices (cf. the estimates in \cite{5}). This
  array of intersecting maximal subsimplices has the natural structure
  of a {\em simplicial complex} with the smaller simplices as faces of
  the maximal ones, viz. the cliques (as to this notion see any
  textbook on algebraic topology like e.g.  \cite{Hopf}, cf. also
  section 3.2 of \cite{3}). If we represent this simplicial complex by
  a new (clique-) graph with only the maximal simplices occurring as
  meta-nodes we loose, on the other side, some information, as we do
  not keep track of, to give an example, situations where, say, three lumps or cliques
  have a common overlap. This situation is not distinguished from the
  scenario where only each pair of the triple has a common overlap.
\end{ob}
\begin{bem} On the other hand, it may be physically advantageous to
  loose microscopic information in a controlled way. We will show
  below that this possibility to consider the network as a simplicial
  complex stops at this first level, the coarser picture of a clique
  graph, however, allows on the other side for a {\em geometric
    renormalization group procedure} as we can repeat this process on
  each level, thus creating a whole tower of such metagraphs, one
  lying over the other, leading in the end to webs which resemble more
  and more our ordinary space-time.
\end{bem}

We now have to make the physically motivated assumption that our
network of lumps, which is assumed to be in the \tit{phase} $QX/ST$,
is not fluctuating too wildly, put differently, that the effective
dynamics, induced by the underlying microscopic dynamics defined
above, leaves the individual cliques, $C_i$, sufficiently stable. We
hence assume that they do not change their form too much from clocktime
step to clocktime step, so that we are able to keep track of the
\tit{history} of the individual cliques over an appreciable amount of
time steps.
\begin{equation}C_i(t)\cap C_i(t+1)\approx C_i(t)\end{equation}
This means that only a small portion of nodes enters or leaves each of
the cliques in every time step. We make this assumption so as to be
able to perform some sort of \tit{assemble averages} over fluctuating but
individual (that is, labelled) cliques in the next section and identify them with
\tit{fuzzy lumps}.
\section{\label{Fuzzy}Cliques as Fuzzy Sets}
We now concentrate on the (clock-time) evolution of an arbitrary
(\tit{generic}) but fixed clique, denoting its time sequence by
\begin{equation}C(t_0),C(t_1),\ldots,C(t_N)\end{equation}
We argued above that the clique under observation changes its shape
and size only mildly from time step to time step. As it is frequently
done in physics, we want to replace the above time series of a generic
clique by an ensemble. To this end we define
\begin{equation}\overline{C}:=\bigcup C(t_i)\quad,\quad
  \underline{C}:=\bigcap C(t_i)\end{equation}
We now define a so-called \tit{membership function}, $m(x)$, for an
arbitrary node, $x$, in the graph, $G$. Traditionally, the usual habit
in \tit{fuzzy set theory} is it, to speak of \tit{fuzzy sets}, but
actually they do exist only via their membership
functions (see below). That is, we relate to each cluster of sharp
cliques, defined above, a so-called \tit{fuzzy clique}, defined by its
membership function. We denote the \tit{fuzzy clique}, belonging to the above
ensemble of sharp cliques by $\widetilde{C}$ and define its membership
function, $ m_{\widetilde{C}}$, as follows:
\begin{defi}[Fuzzy Clique]
For each node, $x$, we define
\begin{equation}m_{\widetilde{C}}:=\begin{cases}0 & x\not\in
    \overline{C}\\
0\le p\le 1 & x\in\overline{C}\\
1 & x\in\underline{C}
\end{cases}
\end{equation}
with $p$ denoting the relative frequency of occurrence of node $x$ in
$\overset{.}{\bigcup}\,C(t_i)$,\,$\overset{.}{\bigcup}$ the
disjoint union. Viz.
\begin{equation}p:= N(x)/N\end{equation}
$N(x)$ the number of occurrences of $x$ in $\overset{.}{\bigcup}\,C(t_i)$.
\end{defi}
Remark: We want to remind the reader of the assumptions we have made
above. Our fuzzy clique or fuzzy lump is a set with the individual
nodes carrying a certain weight, viz. its degree of membership with
respect to the fuzzy clique, $\widetilde{C}$, which, on its side, is
given by the corresponding membership function,
$m_{\widetilde{C}}(x)$. Note that the underlying philosophy is quite
modern as it replaces an underlying space (a set of points) by a class
of functions on the space. A similar philosophy is e.g. hold in
\tit{non-commutative geometry} and related fields.\vspace{0.3cm}

We should add at this place some remarks of precaution. We studied in
quite some detail in \cite{5} the statistical distribution of the
number and size of occurring cliques in, say, a \tit{random
  graph}. It turns out that most of the cliques have a typical
(generic) size. On the other hand there exists a (possibly small)
fraction of degenerated, i.e. very small ones. These small ones may
even happen to vanish and/or emerge in the envisaged clock-time
interval. That is, we expect our general picture to be true modulo
some stochastic noise, which we choose to neglect at the moment.
\begin{conclusion}Under the assumption, being made, we can, in a
  certain approximation, replace the clique graph introduced above by
  a net of overlapping fuzzy lumps. Mathematically this net is
  implemented by the corresponding class of membership functions,
  $\{m_i(x)\}$, with $i$ labelling the fuzzy lumps and where the
  overlap of cliques is now encoded in the overlap of the supports of
  the functions, $m_i(x)$, viz.
\begin{equation}x\in\widetilde{C_i}\quad\text{if}\quad
  m_i(x)>0\end{equation}
\end{conclusion}

We see from the above that, to put it briefly, fuzzy-set theory
consists of replacing the ground set, $\mcal{P}(X)$, $X$ some space of
points, $\mcal{P}(X)$ the set of subsets of $X$, and the respective
operations on it, by the set of functions, $I^X$, $I$ the unit
interval, and corresponding operations on it. $\mcal{P}(X)$ can be made into
a \tit{Boolean algebra} or a \tit{lattice} with the help of the
operations $\cap,\cup$ or $\cap$ as multiplication and the symmetric
difference
\begin{equation}\Delta(A,B):=(A\setminus B)\cup (B\setminus
  A)\end{equation}
as addition.

Corresponding relations can be given for fuzzy sets, but there are a
lot of different possibilities to implement or mimik the above
set-theoretic operations. The perhaps most straightforward  ones are
\begin{equation}\cap\rightarrow
  \min(\widetilde{A}(x),\widetilde{B}(x))\quad
  \cup\rightarrow\max(\widetilde{A}(x),\widetilde{B}(x))\end{equation}
where, for brevity, we use from now on the same symbol for the fuzzy set and its
membership function. Some more remarks on these matters can be found
in the following section. (A nice introduction to fuzzy set theory on
a mathematically satisfying level is e.g. \cite{Lowen}, including a
huge bibliography. A brief but concise representation can also be
found in \cite{Bronstein}). Some slightly more advanced relations
between fuzzy sets, being of potential use in our geometric enterprise
are the ``degree of being a subset'' and the ``degree of similarity''
between two fuzzy sets. Defining the \tit{size} of the fuzzy set by
\begin{equation}[\widetilde{A}]:=\sum_x
  \widetilde{A}(x)\quad\text{or}\quad \int
  \widetilde{A}(x)d^nx\end{equation}
(existence of the rhs being assumed), we define e.g. the degree of
being a subset as
\begin{equation}\operatorname{sub}(\widetilde{B};\widetilde{A}):=[\widetilde{A}\wedge\widetilde{B}]/[\widetilde{A}]\end{equation}
and the degree of similarity of two sets by
\begin{equation}
  \operatorname{sim}(\widetilde{A},\widetilde{B}):=[\widetilde{A}\wedge\widetilde{B}]/[\widetilde{A}\vee\widetilde{B}]\end{equation}
with $\wedge,\vee$ extending the notions of $\cap,\cup$; one may take
e.g. the above $(\min,\max)$-implementations.

The above binary relations, given e.g. by $(\min,\max)$, fullfil the
criteria of so-called $(t,s)$-norms (see the next section). There are
other such functions in use, which have perhaps nicer properties. One
is the so-called \tit{Lukasiewicz-(t,s)-norm}:
\begin{equation}t_L(\widetilde{A},\widetilde{B}):=\max\{0,\widetilde{A}+\widetilde{B}-1\}\quad s_L(\widetilde{A},\widetilde{B}):=\min\{1,\widetilde{A}+\widetilde{B}\}\end{equation}
leading to $\wedge_L,\vee_L$ (for more details see
e.g. \cite{Demant}). The advantage of the Lukasiewicz-operations is
that they induce a metric on fuzzy-sets. We state without proof:
\begin{ob}$1-\operatorname{sim}_L(\widetilde{A},\widetilde{B})$
  defines a metric on fuzzy-sets.
\end{ob}
\begin{bem}We rediscover exactly this $t$-norm in section
  (\ref{metric}) in quite a different context ($E$-spaces).
\end{bem}
As to this observation, note that for ordinary (finite) sets,
$\Delta(A,B)$ induces also a metric.
\begin{ob}[Hamming-distance]
$|\Delta(A,B)|$ defines a metric on finite sets.
\end{ob}
Proof: The only non-trivial property is the \tit{triangle
  inequality}. As a direct approach is perhaps a little bit tedious,
we give a perhaps more pedagogical proof using graph theory. Subsets
of a finite set, $X$, can be represented by functions
$f\in\{0,1\}^X$. These functions, on the other side, can be
represented as the vertices of a hypercube,
$Q^{|X|}$. $|\Delta(A,B)|$ is now the minimal length of a path between
$f_A,f_B$ on $Q^{|X|}$, i.e., the minimal number of flips of $(0,1)$
along a path connecting $f_A$ and $f_B$. This distance on graphs is
however a metric.\bewende     \vspace{0.3cm}

After this short digression we want to return to our original
enterprise, i.e. to relate our approach to the
\tit{probabilistic-metric}-approach, mentioned above. We started from an
underlying graph, $G(t)$, and a superimposed clique-graph, $G_{cl}(t)$, both carrying a
natural distance function, i.e.
\begin{equation}\label{metric1}  d(n_i,n_k)\quad , \quad d_C(C_i,C_k)\end{equation}
that is, the minimal length of a path, connecting the given nodes
(cf. \cite{1} to \cite{5}). Keeping the labelled nodes or cliques
(supernodes) fixed, both these distances fluctuate in the course of
clock-time evolution as bonds are switched on or off according to one
of the microscopic dynamical laws, given above. The clique-metric will
fluctuate since the cliques change their shape and size, viz. also
their degree of overlap. On the other side, its fluctuations are
supposed to be less erratic as a non-void overlap means usually
that several nodes and bonds are involved, hence clique-distance
should be more stable.

When we switch from this dynamical picture of a time-dependent graph,
$G(t)$, to the ensemble picture of fuzzy cliques or lumps, our point
of view changes to a static but, on the other side, probabilistic one.
This latter point of view is more in the spirit of the framework of
Menger et al. That is, the structure of the space under study is no
longer time dependent while its largely hidden dynamics is now encoded
in various probabilistic notions like e.g. a random metric (which we
would like to derive from some underlying principles). We will make
the connection to this other framework after the following sections,
introducing some technical material about fuzzy-set theory and
probabilistic metric spaces. Furthermore we plan to relate in this
final section our approach to the complex of ideas developed by e.g.
Sorkin et al (see e.g. \cite{Bombelli} or \cite{Sorkin}; we recommend
in particular the latter reference as a thoughtful introduction into
this particular bundle of ideas).

\section{Some Concepts from Fuzzy Set Theory}
We give in this section a brief introduction into the ideas of fuzzy
set theory. Let $A$ be a fuzzy set on $X$. Then by definition $A(x)$
is interpreted as the degree to which $x$ belongs to $A$. The fuzzy
operations like fuzzy unions, intersections and complements are
certain (context dependent) generalizations of the corresponding
classical set operations. Let e.g. $B$ be another fuzzy set and $\bar{A}$ be the complement of
$A$. A particular variant of such fuzzy operations are the following:

\begin{equation}
\bar{A}(x) = 1 - A(x)
\end{equation}

\begin{equation}
(A \cap B)(x) = min[A(x),B(x)]
\end{equation}
and
\begin{equation}
(A \cup B)(x) = max[A(x),B(x)]
\end{equation}

In the literature, fuzzy intersections and unions are in general
defined via so-called \tit{t-norms (triangular norms)} and
\tit{t-conorms} respectively. However, it should be noted that such
fuzzy opeartions are by no means unique. Different choices of such
\tit{t-norms} may be appropriate to represent these operations in
varying contexts and hence they are known as context dependent
operations. Essentially fuzzy set theory provides us with an intutive
notion of uncertainty.  Subsequent to the development of fuzzy set
theory, \tit{fuzzy measure theory} was developed (see e.g. \cite{WangKlir}).  The concept of fuzzy measure theory is conceptually an
important step in understanding the foundational issues related to
fuzzy set theory. It provides a broader framework which allows to
introduce something like \tit{possibility theory}.

Crucial in this field as well as in the field of probabilistic metric
spaces is the above mentioned notion of \tit{t-norm} (more about the
history of this important concept can be found in the book of
Schweizer and Sklar, \cite{Schweizer1}).
The (generalized) intersection of two fuzzy sets $A$ and $B$ is defined by a binary operation on the
unit interval as
\begin{equation}t: [0,1]\times [0,1] \rightarrow [0,1]\end{equation}
\begin{defi}[T-Norm]
\begin{equation} t(a,1) = a\quad\text{(boundary condition)}\end{equation}
\begin{equation} b\leq d\;\text{implies}\; t(a,b) \leq t(a,d)\quad\text{(monotonicity)}\end{equation}
\begin{equation} t(a,b) = t(b,d)\quad\text{(commutativity)}\end{equation}
\begin{equation} t(a,t(b,d)) = t(t(a,b),d)\quad\text{(associativity)}\end{equation}
\end{defi}

It is instructive to compare ordinary probability theory with
\tit{possibility theory} and in particular, probability theory with
fuzzy set theory. In some sense a fuzzy measure is the dual concept to
the concept of fuzzy sets. It encodes the possibility of a given fixed
(fuzzy) object to belong to the respective sets of $\mcal{C}$ (cf. also \cite{Gottwald}).
\begin{defi}
Given a set $X$ and a nonempty family $\mcal{C}$ of subsets of $X$,
which, for convenience, we take to be a \tit{$\sigma$-algebra}. A fuzzy
measure on  $(X,\mcal{C})$ is a function
$g : \mcal{C} \rightarrow [0,1]$
that satisfies the following requirements :
\begin{enumerate}

\item  $g(\emptyset) = 0\; \text{and}\; g(X) = 1 $ (boundary requirements).

\item  for all $A,B \in \mcal{C} $, if $A \subseteq B$, then $g(A) \leq g(B)$ (monotonicity).

\item  for any increasing sequence, $A_1 \subset A_2 \subset \cdots $
  in $\mcal{C}$ so that $\bigcup_{i=1}^\infty A_i\in \mcal{C}$ we have
 $\lim_{i\to\infty} g(A_i) = g(\bigcup_{i=1}^\infty A_i)$ (continuity from below)

 \item  for any decreasing sequence $A_1 \supset A_2 \supset \cdots $
   in $\mcal{C}$, $\lim_{i\to\infty} g(A_i) = g(\bigcap_{i=1}^\infty)$ (continuity from above).
\end{enumerate}
\end{defi}

In case of a probability measure we have for $A\cap B=\emptyset$:
\begin{equation}P(A \cup B) = P(A) + P(B)\end{equation}
From this one sees that ordinary probability measures are a true
subclass of possibility or fuzzy measures. As we do not use these more
advanced concepts at the moment, we refrain from giving more details
which can be found in the mentioned literature.

\section{\label{Random} Concepts from the Theory of Probabilistic
  Metric Spaces}
In the statistical geometry, developed by Menger et
al, points are no longer considered as the elementary building
blocks. In some sense lumps play now their role as primordial, not
further resolvable elements. Various concepts of a probabilistic
nature are introduced which allow to quantify varying degrees of
\tit{(in)distinguishability} of objects.  In this way Menger solved
Poincare's dilemma of having, on the one side, a \tit{transitive}
mathematical and, on the other side, a possible intransitive physical relation
of equality (cf. \cite{Poincare} and further remarks in section (\ref
{distance}). In this geometrical framework we have two basic ingredients:
\begin{enumerate}
\item The concept of \tit{hazy} or \tit{fuzzy lumps}
\item  The ``randomisation'' of various geometrical or metrical concepts
\end{enumerate}

Frechet (\cite{Frechet}) gave an abstract formulation of the notion of distance
in 1906. Hausdorff (\cite{Hausdorff}) proposed the name \tit{metric space} and introduced the function $d$ that assigns a nonnegetive real number $d(p,q)$
( the distance between p and q) to every pair $(p,q)$ of elements (points) of
the set $S$. Its properties are
\begin{gather}d(p,q)=0\leftrightarrow p=q \\
d(p,q)=d(q,p) \\
d(p,r)\leq d(p,q)+d(q,r)
\end{gather}
In 1942 Menger (\cite{Menger}), guided by the experimental situation
in the natural sciences, proposed to replace the ``deterministic''
function $d(p,q)$ by a more probabilistic concept. He introduced the
probability distribution, $F_{pq}$, whose value $F_{pq}(x)$
for any real number $x$, is interpreted as the probability that the
distance between $p$ and $q$ is less than $x$.
\begin{bem}We assume that $F_{pq}$ is continuous from the left. On the
  other side, we may nevertheless encounter the situation that
  $F_{pq}^+(0)>0$.  This means that there may be a non-vanishing
  probability for two different lumps to have a vanishing distance.
  This particular point is discussed by Menger in \cite{Menger2} and
  implements Poincare's observation of a possible non-transitive
  behavior.
\end{bem}
 Since probabilities can neither be negative nor be greater
  than $1$, we have
\begin{equation}
0 \leq F_{pq} \leq 1
\end{equation}
for any real $x$.
Menger defined a \tit{statistical metric space} as a set $S$ with an associated set
of probability distribution functions $F_{pq}$ which satisfy the following conditions.
\begin{equation}
F_{pq}(0) = 0
\end{equation}
\begin{equation}
\text{If}\; p = q\;\text{then}\;F_{pq}(x) = 1\;\text{for all}\;x >0
\end{equation}
\begin{equation}
\text{If}\;p \neq q\;\text {then}\;F_{pq}(x) < 1\;\text{for some}\;x >0
\end{equation}
and
\begin{equation}
F_{pq}(x + y) \geq T(F_{pr}(x), F_{qr}(y))
\end{equation}
for all $p,q,r$ in $S$ and all real numbers $x,y$.
Here $T$ is a function from the closed unit square $[0,1]\times [0,1]$ into the
closed unit interval $[0,1]$ (called a \tit{triangular function} or
\tit{triangular norm}, see below).

One may view these probabilistic metric spaces as being derived from
an underlying fully probabilistic model system, living on a
probability space and with the physical observables being random
variables. Spacek \cite{Spacek} was the first to look at the subject
from this point of view.  He proposed the name \tit{random metric
  space} instead of probabilistic metric space and discussed the relationship
between these two notions. Stevens
\cite{Steven} in his doctoral dissertation tried to modify Spacek's
approach. The main idea behind Steven's approach lies in the fact that
one has a set $S$ and a collection ${\it P}$ of measuring rods. One
chooses a measuring rod $d$ from ${\it P}$ at random to
measure the distance between two given points, $p$ and $q$, of
$S$. With the help of this idea Stevens defined the distribution function $F_{pq}$ and
showed that the metrically generated space so obtained is a Menger
space.

Menger, on the other side, started with a probability distribution
function instead of random variables. This is related to the fact that
the outcome of any series of measurements of the values of a
nondeterministic quantity is a distribution function and the
probability space may be unobservable in principle. This point of view
has its roots in the positivistic philosophy (cf. e.g. \cite{Menger2})
and is in line with a possible nonclassical behaviour.

Sherwood \cite{Sherwood} approached the problem from a different point
of view.  Following the concept of distribution-generated spaces as introduced by
Schweizer and Sklar (\cite{Schweizer}), Sherwood proposed the concept
of $E$-space. In an $E$-space, the points are functions from a probability
space $(\Omega, \mcal{A},P)$ into a metric space $(M,d)$. For each
pair, $(p,q)$, of functions in the space, the function $d(p,q)$, defined as
\begin{equation}(d(p,q))(\omega) = d(p(\omega),q(\omega))
\end{equation}
for all $\omega$ in $\Omega$, is a random variable on $(\Omega,
\mcal{A},P)$. The function $F_{pq}$ is then the distribution function
of this random variable, i.e. we have
\begin{equation}F_{pq}(x) = P{((d(p,q))(\omega) < x)} \end{equation}
In this way, $F_{pq}$ can be regarded  as the probability that the
distance between $p$ and $q$ is less than $x$. Sherwood showed that
every E-space is a Menger space (cf. the subsection (\ref{metric}) below).

Another subclass of probabilistic metric spaces are the so-called
distribution generated spaces. The main idea is roughly as follows:
Let $S$ be a set. With each point $p$ of $S$ associate an n-dimensional
distribution function $G_p$ and with each pair, $(p,q)$, a
2n-distribution function $H_{pq}$ so that it holds:
\begin{equation}
H_{pq}(\vec u, \vec v=(\infty,.......,\infty)) = G_p(\vec u) \end{equation}
\begin{equation}H_{pq}(\vec u=(\infty,.......,\infty),\vec v) = G_q(\vec v)
\end{equation}
for any $\vec u = (u_1,.....,u_n)$ and $\vec v = (v_1,.......,v_n)$ in
$R^n$.

Let now $Z(x)$ be a cylinder in $R^n$ with
\begin{equation}Z(x):= \left\{(\vec u, \vec v) \in R^{2n}\,;\, |\vec u - \vec v|
< x \right\}\end{equation}
for any $x>0$. This defines a distance distribution function,
$F_{pq}$, via
\begin{equation}
F_{pq}(x) = \int_{Z(x)} dH_{pq} =:P_{H_{pq}}(Z(x))
\end{equation}
The technical details can be found in e.g. \cite{Schweizer1}. Note
that, in general, the pairs of $S$ need not be independently distributed.

In one interpretation one may view the elements of $S$ as
``particles''. Then for any Borel set $A$ in $R^n$, the integral
$\int_A dG_p$ can be interpreted as the probability that the particle
$p$ can be found in the
set $A$ and $F_{pq}(x)$ as the probability that the distance between
the particles $q$ and $p$ is less than $x$. This then yields another
type of probabilistic metric space.

A subclass consists of models where the members in $S$ behave
independently of each other. Such spaces are called
\tit{cloud-spaces(C-spaces)}. A function
$g$ from $R^n$ into $R^+$ is an n-dimensional probability density if the function
$G$ defined on $R^n$ by
\begin{equation}
G(\vec u) = \int_{((-\infty,......,-\infty),\vec u))} g(\vec v) d\vec v
\end{equation}
is an n-dimensional distribution function. If $p$ is a point in a
distribution-generated space over $R^n$ such that $G_p$ is absolutely
continuous, then the corresponding density $g_p$ of $G_p$ may be visualized as a
"cloud" in $R^n$ - a cloud whose density at any point of $R^n$
measures the relative likelihood of finding the particle $p$ in the
vicinity of that point.

Another, equally natural interpretation (which is perhaps more in line
with our concept of \tit{fuzzy lumps}; see section (\ref{Fuzzy})) is
to visualize the space as an aggregate of \tit{clouds} or \tit{fuzzy
  points}. To make life simple we may assume that to all points, $p$,
is associated a vector, $\vec{c}_p$ in $R^n$ so that
\begin{equation}g_p(\vec{u})=g(\vec{u}-\vec{c}_p)\end{equation}
with $g$ sperically symmetric. We may then replace the points by
clouds in $R^n$ and get a notion of \tit{homogeneity}. This model is
perhaps the most straightforward extension of ordinary (Euclidean)
space and resembles our network model of fuzzy lumps (viewed as fuzzy
sets) if we chose to embed it into ordinary $R^n$.

It can be shown that by taking the convolution product of
$g(\vec{u}-\vec{c}_p)$ and $g(\vec{v}-\vec{c}_q)$, one can generate a
particular type of random metric space. By applying then a further
amount of probabilistic machinery one can create what is called
\tit{Frechet-Minkowski-metrics}, $d_{\beta}$ on the original space
$S$, which may e.g. be classical Euclidean space. These metrics are
related with and derived from the underlying random metric space
(cloud space). The to some extent intricate calculations can be found
in \cite{Schweizer1}, chapter 10.

These \tit{Frechet-Minkowski metrics}, $d_{\beta}$, associated with
(semihomogeneous) $C$-spaces over $R^n$ have a remarkable structure.
At small distances this metric is non-Euclidean and the distance
between two distinct points $p,q$ is bigger than a fixed positive
constant, associated with $g(\vec{u})$. On the other hand, it becomes
euclidean in the  asymptotic region.  It appears from the above picture
that if we consider a $C$-space as a space of clouds (which may move
around), the ``haziness'' of the distance between $p$ and $q$, which is
predominant when the clouds are close together, becomes more and more
insignificant when the clouds are sufficiently far apart. In this
sense \tit{Frechet-Minkowski metrics} become asymptotically
euclidean.

\section{\label{distance}Random Metrics on our Networks of Fuzzy Lumps}
In this section we want to relate the various complexes of concepts
and techniques, developed or presented above, with each other. That
is, we view on the one side the (dynamically) fluctuating cliques as
smeared out lumps, viz. as fuzzy sets. On the other side, we want to
construct so-called \tit{probabilistic} or \tit{random metric spaces}
over this space of fuzzy lumps.

Note that this task is however not entirely straightforward as within
the mathematical framework of the latter approach the underlying space
is usually treated as a more or less structureless space of simple
points with the underlying possible causes of the fluctuations in the
metrical distance usually not being openly discussed. This has however
to be done in the concretely given model systems.

We should ad the remark that there are also different scientific
philosophies behind these various points of view (cf. \cite{Schweizer1}
p.17). There exists a tradition to keep the (or a) underlying concrete
\tit{probability space} out of the game and regard is as secondary,
while concentrating on the class of (phenomenologically given)
\tit{distribution functions}.

On the other side, a concretely given (and, what is even better,
physically motivated) microscopic probabilistic ground space provides
(among other things) a common and unifying reference frame. In other
words, we have a canonical deterministic metric on the underlying
dynamical graph, $G(t)$ or its \tit{clique graph}, $G_{cl}(t)$
(cf. (\ref{metric1})). We have argued that our cliques or lumps are
fluctuating as a result of the imposed dynamical laws (discussed in
section 2). From this input we have to infer the concept of a
\tit{probabilistic distance} between two given labelled (fuzzy) lumps.
We divide our section into two subsections. In the first subsection we
introduce distance concepts on the space of overlapping (static) fuzzy
lumps as they are given in the context of \tit{fuzzy set theory} as
discussed in section 4. In this context fluctuations are only
implicitly included and the model theory has (one may say) the status
of kind of a \tit{mean field theory}. On the other side, the metrics,
we introduce in the following subsection are related in spirit to
distance concepts introduced by Menger in his interesting essay
\cite{Menger2}. More fundamental are in our view the concepts we
develop in the other subsection. We construct an explicit probability
space, introduce metrical distance as an explicitly given random
variable, find the corresponding \tit{triangular norm} and show that
the space we get is a model of a so-called \tit{E-space}, introduced
in section \ref{Random}. These E-spaces are a subclass of Mengers
probabilistic metric spaces.
\subsection{Metrics on Fuzzy Lumps, the Mean Field Picture}
The perhaps most immediate method to impose some metric distance on
our space of fuzzy lumps is the following one. In a first step we
define a distance concept for the immediate (``infinitesinal'')
neighborhood of a given lump and then proceed (for finite distances)
by a \tit{chain method}, viz. paste infinitesimal distances together.

We start with two fuzzy lumps, $\widetilde{C}_0,\widetilde{C}_1$,
which overlap as fuzzy sets. We hence have (see section 4):
\begin{equation}0\leq
  p(\widetilde{C}_0,\widetilde{C}_1):=1-\operatorname{sim}(\widetilde{C}_0,\widetilde{C}_1)<1\end{equation}
We have now several possibilities to proceed. We can e.g. fix a
certain scale resolution and state that two fuzzy lumps are
indistinguishable  if
\begin{equation}p(\widetilde{C}_0,\widetilde{C}_1)\leq\varepsilon\end{equation}
This is in accord with already classical ideas of Poincar\'e,
developed in \cite{Poincare}, p.31f. Put differently, in the physical
world we may have a \tit{non-transitive} infinitesimal distance
concept or, more generally, a non-transitive concept of \tit{identity}
so that
\begin{equation}A=B\;,\; B=C\;,\; A\neq C\end{equation}
or
\begin{equation}d(A,B)=0\;,\; d(B,C)=0\;,\; d(A,C)\neq 0\end{equation}
\begin{defi}[Distance, Variant 1] Denoting by $\gamma$ a path
  connecting $\widetilde{C}_i$ and $\widetilde{C}_j$, which consists
  of a chain s.t.
\begin{equation}\widetilde{C}_i=:\widetilde{C}_{k_0}\;,\;\widetilde{C}_j=:\widetilde{C}_{k_n}\end{equation}
and
\begin{equation}p(\widetilde{C}_{k_l},\widetilde{C}_{k_{l+1}})\leq
  \varepsilon\end{equation}
we define
\begin{equation}\operatorname{dist}_{\varepsilon}(\widetilde{C}_i,\widetilde{C}_j):=\inf_{\gamma}n\end{equation}
($\operatorname{dist}_{\varepsilon}:=\infty$ if there is no such
chain).
\end{defi}

It can be easily shown that this definition fullfils the axioms of a
metric. Note however that the validity of the \tit{triangle
  inequality} is enforced by attributing the distance `one' to two
``indistinguishable'' lumps. This is a little bit unpleasant but
inescapable if we want to base our macroscopic distance concept on the
concatenation of ``infinitesimal'' distances. On the other side, this
kind of distance is probabilistic in a very restricted sense and
yields only distances which are multiples of a fixed given distance
element.

We can improve the situation by taking as infinitesimal distance
elements the above measures of overlap,$
p(\widetilde{C}_{k_l},\widetilde{C}_{k_{l+1}})$, and define
\begin{defi}[Distance, Variant 2]
\begin{equation}d(\widetilde{C}_i,\widetilde{C}_j):=\inf_{\gamma}\sum
  p(\widetilde{C}_{k_l},\widetilde{C}_{k_{l+1}})\end{equation}
\end{defi}
Note that in contrast to the preceding definition now the infimum need
not be taken at a path of minimal canonical length. Quite the
contrary, in exceptional situations the canonical length of a path
leading to a very short distance may be quite large, whereas this is
not expected to be the ordinary situation. Anyway, the second distance
concept is a little bit more probabilistic.
\subsection{\label{metric}The Space of Cliques as a Probability Space and the
  Connection with Random Metric Spaces}
We repeat briefly what we have said in preceding sections. We have
assumed that we have a network of a more or less fixed number of,
however, fluctuating cliques. That is, their degree of overlap may
vary in the course of time. At each discrete time step during a
sufficiently long time interval, $I$, (so that we can exploit
probabilistic concepts in, at least, a certain approximation) we have
a definite array of overlapping cliques, $G_{cl}(t)$. As ground set
for our probability space we take this set, i.e.
\begin{equation}X:=\{G_{cl}(t),t\in I,|I|=N,\;\text{$N$ some sufficiently
    large number}\}\end{equation}
The simplest assumption is to assume that all configurations have the
same probability, $P(G_{cl}(t)):=1/N$, say.
\begin{bem}On physical grounds a more sophisticated probability
  density may suggest itself.Our results, derived below, are however
  independent of the particular choice of $P$.
\end{bem}
\begin{ob}The above defines a discrete probability space,$X$, on which
  random variables can be introduced, their expectation value being
  denoted by $<\;>$.
\end{ob}
\begin{bsp}We previously introduced the membership function,
  $\widetilde{C}_i(x)$, of a fuzzy lump, $\widetilde{C}_i$. Defining
  the elementary random variable ($x$ a node of the primordial
  underlying graph)
\begin{equation}x_{C_i}(G_{cl}(t))=\begin{cases}1 & x\in C_i(t) \\
0 & \text{else}\end{cases}\end{equation}
we have
\begin{equation}\widetilde{C}_i(x)=\langle
  x_{C_i}\rangle=:x(\widetilde{C}_i)\end{equation}
(cf. the section about fuzzy lumps).
\end{bsp}

In the same sense we can define the distance between two fixed lumps,
$C_i,C_j$, as a random variable on our probability space, taking the
values $d_{(C_i,C_j)}(G_{cl}(t))$. In contrast to our ordinary fuzzy
lumps, all these random variables do fluctuate, the fluctuations given
by
\begin{equation}\langle(x_{C_i}-\langle
  x_{C_i}\rangle)^2\rangle\quad\text{or}\quad \langle(d-\langle d\rangle)^2\rangle\end{equation}
For simplicity reasons we write from now on the metric as $d(p,q)$,
$p,q$ denoting some cliques, and with the understanding that it
represents the above random variable.

If we want to relate our approach to the probabilistic metric
framework laid out in e.g. \cite{Schweizer1}, we have to inspect the
\tit{transitivity properties} of our distance concept.
\begin{ob}With $p,q,r$ three points (cliques), we have for the
  respective random variables
\begin{equation}d_{p,r}\leq d_{p,q}+d_{q,r}\end{equation}
as they are evaluated on a fixed network, $G_{cl}(t)$, where the
ordinary triangle inequality of the canonical graph distance
holds. Hence
\begin{equation}\langle d_{p,r}\rangle\leq \langle
  d_{p,q\rangle}+\langle d_{q,r}\rangle \end{equation}
\end{ob}

On the other side, probabilistic metric spaces are defined via the
probability distributions, $F_{pq}(x)$, denoting the probability that
the distance between $p,q$ is smaller than $x$. The ordinary triangle
inequality is replaced in this framework by a more complicated
estimate (cf. section \ref{Random}):
\begin{equation}F_{pr}(x+y)\geq T(F_{pq},F_{qr})\end{equation}
with $T$ a \tit{triangular norm}.

Our task consists now in determining this function $T$ within our own
stochastic framework. To this end we employ the following formula
which holds for probability measures (i.e. $P(X)=1$).
\begin{lemma}With $A,B$ measurable sets in $X$ it holds
\begin{equation}P(A\cap B)\geq P(A)+P(B)-1\end{equation}
\end{lemma}
Proof: This result has also been employed in \cite{Schweizer1},
p.26. For completeness sake we repeat it here. We have
\begin{multline}P(A\cap B)= P(A)+P(B)-P(A\cup B)\geq
  P(A)+P(B)-P(X)\\=P(A)+P(B)-1\end{multline}
and therefore
\begin{equation}P(A\cap B)\geq\max(P(A)+P(B)-1,0)\end{equation}
This proves the lemma.\bewende

We have by definition:
\begin{equation}F_{pr}(x+y)=P(d_{pr}<x+y)\end{equation}
which, for the simplest choice of $P$, is just the fraction of configurations,
$G_{cl}(t)$, with the canonical graph metric fulfilling this
inequality. As the triangle inequality holds for the canonical graph
metric we have
\begin{equation}P(d_{pq}<x,d_{qr}<y)\leq P(d_{pr}<x+y)\end{equation}
and hence, with the help of the above lemma:
\begin{equation}F_{pr}(x+y)=P(d_{pr}<x+y)\geq\max(F_{pq}(x)+F_{qr}(y)-1,0)\end{equation}
\begin{ob}We proved that our random metric, introduced above, fulfills
  a triangular inequality
\begin{equation}F_{pr}(x+y)\geq W(F_{pq}(x),F_{qr}(y))\end{equation}
with $W$ the triangular norm
\begin{equation}W(a,b):=\max(a+b-1,0)\end{equation}
Hence our above probability space is a Menger space.
\end{ob}
(That $W$ is in fact a triangular norm is easy to show and is also
employed in e.g. \cite{Schweizer1}).

A special class of probabilistic metric spaces are the so-called
\tit{E-spaces} (cf. the section \ref{Random} and \cite{Schweizer1},
chapt. 9). To complete our analysis we can reframe our model so that
it becomes such an \tit{E-space}. The ground
space, $X$, is our probability space. The metric target space, $M$, is
a new \tit{space-time supergraph}, defined as follows:
\begin{defi}[Space-Time Picture of Clique Graph]
\begin{equation}M:=\bigcup_{t\in I}G_{cl}(t)\end{equation}
with the understanding that exactly the nodes $C_i(t)$ and $C_i(t+1)$,
$t,t+1\in I$, are connected by additional bonds. That is, $C_i(t)$
describes the orbit of the fixed but fluctuating clique, $C_i$, in the
course of time.
\end{defi}
We can now relate bijectively the cliques, $C_i$, and certain
functions from $X$ to $M$.
\begin{ob}With the class of functions, $C_i$,
\begin{equation}C_i(G_{cl}(t)):=C_i(t)\in M\end{equation}
and their distance
\begin{equation}d_{pq}(G_{cl}(t)):=d(C_p(t),C_q(t))\end{equation}
the above probability space can be considered as an $E$-space.
\end{ob}

\end{document}